\documentclass[prd,article,11pt]{revtex4}%
\usepackage{amsmath, amsthm, amsfonts}
\usepackage[spanish,activeacute,english]{babel}
\usepackage{amsmath}
\usepackage{amsfonts}
\usepackage{amssymb}
\usepackage{graphicx}
\usepackage{cancel}
\usepackage{appendix}
\usepackage{graphicx}
\usepackage{vmargin}
\usepackage{makeidx}
\usepackage{float}
\usepackage{subfig}
\usepackage{braket}
\usepackage{dsfont}
\usepackage{hyperref}
\usepackage{color}
\usepackage{amsmath}
\usepackage{graphicx}
\usepackage{epsfig}
\usepackage{dcolumn}
\usepackage{bm}
\usepackage{dcolumn}
\usepackage{amsfonts}
\usepackage{amssymb}%
\providecommand{\U}[1]{\protect\rule{.1in}{.1in}}
\providecommand{\U}[1]{\protect\rule{.1in}{.1in}}

\theoremstyle{definition}

\theoremstyle{remark}

\begin{document}
\title{On the dynamics of the contagious rate under isolation measures}

\author{ Alejandro Cabo Montes de
Oca $^{1}$\thanks{cabo@icimaf.cu} and Nana Cabo Bizet $^{2}$\thanks{nana@fisica.ugto.mx} \medskip}

\affiliation{$^{1}$\textit{Departamento de F\'isica Te\'orica, Instituto de
Cibern\'{e}tica, Matem\'{a}tica y F\'{\i}sica, Calle E, No. 309, Vedado, La
Habana, Cuba }}

\affiliation{$^{2}$ \textit{Departamento de F\'{\i}sica, DCI, Campus Le\'on, Universidad de
Guanjuato, \ CP. 37150, Le\'on, Guanajuato, M\'exico }}

\begin{abstract}
\noindent  The infection dynamics of a population under stationary isolation conditions is modeled.
It is underlined that the stationary character of the isolation
measures can be expected to imply that an effective  SIR model with constant parameters
should describe the infection process. Then, a derivation of this property is
presented, assuming that the statistical fluctuations in the number of
infection and recovered cases are disregarded.  This effective SIR model shows a
reduced population number and a constant $\beta$ parameter. The effects of also including
the retardation between recovery and infection process is also considered. Next, it is  shown
that any solution of the effective SIR also solves the linear problem to which
the SIR equations reduce when the total population is much larger than the number of
the infected cases. Then, it is also argued  that this equivalence follows for a specific
contagious parameter $\beta(t)$ which time dependence is analytically
derived  here. Then, two equivalent predictive calculational methods for the infection dynamics
under stationary isolation measures are proposed. The results represent a solutions for the known and challenging problem of defining the time dependence of the contagion parameter, when  the SIR parameter $N$ is assumed to be the whole population number. Finally, the model is applied to describe the known infection curves for countries
that already had passed the epidemic process under strict  stationary isolation measures.
The cases of Iceland, New Zealand,  Korea and Cuba were considered. Although, non subject to
stationary isolation measures  the cases of U.S.A. and Mexico are also examined due to their interest.
 The results support the argued validity of SIR model including retardation.

\end{abstract}
\maketitle

\newpage
\section{Introduction}

As following from its relevance for the human beings, the investigation of the
covid-19 epidemic had attracted a lot of attention in the actual research
literature \cite{1,2,3,4,5,6,7,8,9,10,11,ours,12,13,DamianNana,harko,14,15,16,17,18,19}. Currently, a large number of
information centers publish the epidemic data which are reported daily for
almost all the counties in the planet \cite{9}, \cite{11}. This situation
furnish to the researchers abundant information about the pandemic.

In a previous work in reference \cite{ours} we had underlined the interest of
considering the dynamics of the contagion parameter $\beta$ of the SIR model
under isolation measures over the populations.  As it is known, this model is
one of the main ones in epidemiology. It has the whole number of habitants
$N_{pop}$ as one of its parameters. Therefore, if the population of the
country $N_{pop}$ is assumed as the $N$ parameter, and the number of infected
cases is very much smaller than $N_{pop},$ the linearity of the equations
enforces that the model parameters should change in time as a consequence of
the imposed isolation measures, if maxima of the infected cases appear. In
that work, we formulated a model in which the initially constant contagion
parameter $\beta$ suddenly drops to a smaller value just after the isolation
measures are installed. Further, it was assumed that these measures should
also continuously reduce the value of the contagion parameter. This was
expected to occur because the infected persons will be confined from to be in
contact with most of the other persons. Thus, they can infect only close
members of the family, and only up to the mean time for which the infected
person remains sick. But, considering the unreal limit of all the habitants
being isolated one from any other, it is clear that the contagion parameter
should become exactly zero after a mean time of occurrence of the affection.
Therefore, we fixed the parameter $\beta$ drastically to zero after a mean
time of duration of the sickness following the imposition of the confinement
rules. This assumption was taken only for countries in which the isolation
measures had been rigorously and stationary maintained,  as in Germany. The
application of this simple model was able to appropriately describe the curve
of infection for this country \cite{ours}.

In the present work, we investigate the infection process as occurring in
countries in which strict isolation measures had been taken, showing a
stationary in time character. That is, measures that remain applied during a
time lapse without serious modifications which could change the contagion
dynamics in the Society.

The work firstly presents a proof that, under the above conditions, the
epidemic should be described by an effective SIR process, having a reduced
population number parameter, as well as a contagion $\beta$ one. The recovering
parameter, can not be expected to be altered by the isolation measures, since
it only depends of the society  structure  and the established action
of the Health System of each country.

Also, the discussion  considers the incorporation of the retardation between the
recovery and the infection of the affected persons. The effect was described
by a new retardation time parameter $\tau.$ This consideration should be taken
into account for describing the epidemic data reported \cite{7}.

In addition it was  took into account  a characteristic property of the covid-19
epidemic. It is the large number of persons which shows the sickness in a mild
form, not allowing them to be detected by the Health Systems. It is then
considered that the observed data of infected persons is a definite fraction
$k$ of the total number of infected persons  \cite{5}. This fraction had been estimated
to have a value nearly to $k=0.2$. This effect is used to define the initial
conditions of the SIR model equations in terms of the observed data.

A second main aim of this study consists in to  discuss the SIR models assuming that the parameter $N$
represents the  total population of the  country and the  number of infected
cases is very much smaller than $N$.   It can underlined that this class of studies had been widely considered
in epidemiology research along the times \cite{14,15,16,17,18,19}. In them, it had been always unclear
how to define the required time dependence of the contagion parameter which becomes required if the epidemic evolutions
show maxima which are very much smaller than the country population.  It should be noted that when  the parameter $N$ is the total population and the number of infected persons is still very much smaller than $N$,  the SIR equations become linear, and not showing any maxima as time follows if the parameters of the model are strictly constant \cite{ours}.
  The present work intends to clarify this situation by  identifying the time dependent contagion parameter
 which should be assumed in order to  define the solution of the linear model describing  the evolution of the epidemic under   stationary isolation conditions.

 For the above defined purpose, it is firstly shown that solutions of the effective SIR equations $S_{e}%
(t),I_{e}(t)$ and $R_{e}(t)$ are also solutions of the linear SIR equations
for large populations and small number of infected persons for special definite time dependent contagion parameter $\beta(t)$. Further, this  time dependent $\beta(t)$  is exactly evaluated  as a function of the  parameters $N_{e}$, $\beta_{e}$ and $\gamma$ of the  effective SIR.  Therefore, the time dependence of  the contagion parameter required for defining the epidemic evolution under isolation conditions of the populations is exactly determined, for the situation in which the parameter $N$ of the SIR model is made equal to the population number.

Finally, the model including retardation is applied to the reported infection
data for various countries. The epidemic data were mainly collected from
reference \cite{9}. The effective SIR equations including retardation were
firstly applied to Iceland and New Zealand. These two countries are known by
their effective isolation measures that were able to finish the epidemic.
After, fixing the initial data for the confirmed, infected and recovered
cases, it was possible to determine the four parameters of the model ($\beta$,
$N$, $\gamma$ and $\tau$). They were chosen in order to define a solution for
the equations which closely resembles the observed data presented in reference
\cite{9} for those two countries. The result confirms that the SIR equations are
closely satisfied by the infection dynamics under stationary isolation
conditions. Further, it was investigated the epidemic infection curve
associated to South Korea. Following the same procedure as for Iceland and New
Zealand, the solution determined by the reported initial data, also closely
approached the data curves for confirmed, infected and recovered cases up to
the 4 March 2020. After this time the solutions deviated form the measured
data. Then, the information on the isolation measures for South Korea was
consulted. With surprise, we found that precisely the 4 March the government
of the country had enforced new drastic isolation measures, which had
disrupted the stationary property of the before taken rules. Therefore, a
clear explanation emerged for the observed deviation of the SIR solution from
the data after the 4 March. Next, the case of Cuba was also studied. The
stationary character of the applied rules was also checked since the data were
also closely matching by a definite solution of the SIR equations including retardation.

It should be underlined that in all the above mentioned countries, non
vanishing retardation between the recovery and infection were needed to
closely approach the observed data. Also, the recovery rate parameters
determined for optimally approaching the observed data were also closely near
the value 0.05.

We applied to model to the case of United States and Mexico. The epidemic had
taken its strongest form in USA at the time of ending the writing of this
work.  In Mexico, a
large country, the epidemic had taken an irregular evolution. Therefore, we decided to explore the possibility of estimating
the infection curves  for both countries. However, it can not
be assured that U.S.A. and Mexico had established well defined stationary and
globally applied isolation measures. Thus, the present exploration should be
only considered as a possibly helpful rough estimation of an important
information for these two strong epidemics.

The work proceeds as follows. In Section 2, it is presented the argue about
that the SIR model with modified parameters should be valid for describing the
stationary isolation infection measurements. It is also introduced the retardation of the
SIR equations, which should be expected to be relevant in comparing the
solutions with the observed data.

Further, initial conditions for the SIR equations appropriately describing the
large number of asymptomatic cases in the covid-19 epidemic are defined in
Section 3.

Next, the Section 4 shows that the effective SIR solutions
also solve the linear SIR equations for small ratio between the number of
infected cases and the total population.

The solutions of the equations are considered in various subsections of
Section 5, in order to compare the predictions of the SIR model with
retardation with the observed infection data in the literature for various
countries.    The conclusions shortly resume the results of the paper and mention possible extensions.

\section{Populations under stationary isolation measures}

Let us assume that the country population $N$ is grouped in families having a
probability distribution of having a specified number of elements in the
family. Just after the given time of imposition of the isolation measures,
there will be a number of infected persons, that will be localized in a number
of families $F_{\inf}$ .  They will be called as infected families. The rest
of families $F_{hea}=F-F_{\inf}$ will be named as healthy ones. But, since the
isolation conditions will be assumed to avoid the contagious between families,
the number $F_{hea}$ should expected to be nearly constant in time, as it will
be also the number of infect ones $F_{\inf}.$ These nearly conserved numbers
can be suspected as defining the new effective population parameter.

Let us separately consider in what follows the evaluations of the rate of
infection of the susceptible persons and the rate of recuperation of the cases.

\subsection{The recuperation rate}

This is the most simple of the quantities to determine, because it is
reasonably to consider it as well described by same equation valid for the SIR
model
\begin{equation}
R(t)=\gamma\text{ }I(t),
\end{equation}
where numerical estimates of the recuperation rate parameter $\gamma$ exist in
the literature. This constant in time rate is expected to be valid because the isolation
conditions should not change the probability of recuperation of a given
infected person as  randomly determined in mean by his physiology and  the
standard treatments of the Health System  in his country.

\subsection{The infection rate}

Let us consider now the infection rate under isolation conditions. For this
purpose define the number of infected persons $n_{i}$ in each infected family
with the subindex $i$, running over $i=1,2,3,....,F_{\inf}.$ The total number
of members of each family will be indicated by $j_{i}$

Consider now a given infected person in his neighborhood of $j_{i}$ familiar
persons and define the probability $p$ that he infects another person during a
day interval (the unit of time assumed). Consider also that in any family $i$
there are $n_{i}^{R}$ recuperated members in addition to the $n_{i}$ infected
ones. In these conditions there will be $(j_{i}-$ $n_{i}-n_{i}^{R})$ remaining
susceptible members.

Then, the probability that a given sick person infects within a day a member
of his family can estimated by the product of the probability for infecting a
person by the number of susceptible persons in the family. Since there are
$n_{i}$ infected persons in the family, the probability should be also
multiplied by $n_{i}$. Then, it results that the change in the number of
susceptible persons $S^{\ast}$ per unit time can be estimated in the form
\begin{equation}
\frac{dS^{\ast}(t)}{dt}=-\sum_{i}n_{i}\text{ }p\text{ }(j_{i}-n_{i}-n_{i}%
^{R}).
\end{equation}
Define now the mean numbers of infected $\overline{n}(t)$ and recovered
$\overline{n}^{R}(t)$ persons, and the mean number of elements in the infected
families $n$ at a given time $t$ as
\begin{align}
\overline{n}(t)  &  =\frac{1}{F_{\inf}}\sum_{i}n_{i}(t)=\frac{1}{F_{\inf}%
}I(t),\label{1}\\
\overline{n}^{R}(t)  &  =\frac{1}{F_{\inf}}\sum_{i}n_{i}^{R}(t)=\frac
{1}{F_{\inf}}R(t),\label{2}\\
n  &  =\frac{1}{F_{\inf}}\sum_{i}j_{i}. \label{nI}%
\end{align}

Then, we can write $\frac{dS^{\ast}(t)}{dt}$ in the form
\begin{align}
\frac{dS^{\ast}(t)}{dt}  &  =-\sum_{i}n_{i}\text{ }p\text{ }(j_{i}-n_{i}%
-n_{i}^{R})\nonumber\\
&  =-F_{\inf}\text{ }n\text{ }p\text{ }\overline{n}(t)+p\sum_{i}(n_{i}%
^{2}\text{ }+n_{i}n_{i}^{R})\nonumber\\
&  =-\text{ }n\text{ }p\text{ }I(t)+p\sum_{i}(n_{i}^{2}\text{ }+n_{i}n_{i}%
^{R}). \label{S}%
\end{align}

The first term in the last line of equation (\ref{S}) is already expressed in
terms of the whole number of infected persons $I(t)$ as in the SIR equations.
However, the second term is a sum of quadratic terms in the numbers of infects
and recuperated persons in the families.

But, it is possible to write for them%
\begin{align}
\sum_{i}n_{i}^{2}  &  =\sum_{i}(n_{i}-\text{ }\overline{n}(t)+\text{
}\overline{n}(t))^{2}\nonumber\\
&  =\sum_{i}(n_{i}-\text{ }\overline{n}(t))^{2}+\text{ }\overline{n}%
(t)^{2}+2\text{ }\overline{n}(t)(n_{i}-\text{ }\overline{n}(t))\nonumber\\
&  =\text{ }\overline{n}(t)^{2}+\sum_{i}(n_{i}-\text{ }\overline{n}(t))^{2},\\
\sum_{i}n_{i}n_{i}^{R}  &  =\sum_{i}(n_{i}(t)-\text{ }\overline{n}(t)+\text{
}\overline{n}(t))(n_{i}^{R}-\text{ }\overline{n}^{R}(t)+\text{ }\overline
{n}^{R}(t))\nonumber\\
&  =\text{ }\overline{n}(t)\overline{n}^{R}(t)+(n_{i}(t)-\text{ }\overline
{n}(t))(n_{i}^{R}(t)-\text{ }\overline{n}^{R}(t))+\nonumber\\
&  (n_{i}(t)-\text{ }\overline{n}(t))\overline{n}^{R}(t)+\overline{n}%
(t)(n_{i}^{R}(t)-\text{ }\overline{n}^{R}(t))\nonumber\\
&  =\overline{n}(t)\overline{n}^{R}(t)+(n_{i}(t)-\text{ }\overline
{n}(t))(n_{i}^{R}(t)-\text{ }\overline{n}^{R}(t)).
\end{align}

We will consider that the number of infected $n_{i}$ and recovered $n_{i}^{R}$
persons in the number of infected families show small deviations from the
respective mean numbers $\overline{n}(t)$ and $\overline{n}^{R}(t)$. That is,
the following relations are obeyed
\begin{align*}
|n_{i}(t)-\text{ }\overline{n}(t)|\text{ }  &  \ll\text{ }\overline{n}(t),\\
|n_{i}^{R}(t)-\text{ }\overline{n}^{R}(t)|\text{ }  &  \ll\text{ }\overline
{n}^{R}(t).
\end{align*}

Therefore, the statistical fluctuations in those number may be disregarded.

  Under these assumption the expression (\ref{S}) can be written only in terms of
the mean numbers of infected and recovered cases as
\begin{align}
\frac{dS^{\ast}(t)}{dt}  &  =-n\text{ }p\text{ }I(t)+pF_{\inf}(\overline
{n}^{2}\text{ }+\overline{n}\text{ }\overline{n}^{R})\\
&  =-n\text{ }p\text{ }I(t)+pF_{\inf}((\frac{I(t)}{F_{\inf}})^{2}\text{
}+\frac{I(t)}{F_{\inf}}\frac{R(t)}{F_{\inf}})\\
&  =-p\text{ }I(t)\text{ }{\Large (}n-\frac{I(t)}{F_{\inf}}-\frac
{R(t)}{F_{\inf}}{\large ).}%
\end{align}

Thus, it is possible to write now the set of equations for the system. After
considering the exact relation%
\begin{equation}
N=S^{\ast}(t)+I(t)+R(t),
\end{equation}
which implies%
\begin{equation}
\frac{dI(t)}{dt}=-\frac{dS^{\ast}(t)}{dt}-\frac{dR(t)}{dt},
\end{equation}
and employing (\ref{1}), (\ref{2}), (\ref{nI}) and (\ref{S}), the set of
equations for modeling the isolation condition of the population can be
written as follows
\begin{align}
\frac{dS^{\ast}(t)}{dt}  &  =-p\text{ }{\Large (}n-\frac{I(t)}{F_{\inf}}%
-\frac{R(t)}{F_{\inf}}{\large )}\text{ }I(t),\label{eq1}\\
\frac{dI(t)}{dt}  &  =p\text{ }{\Large (}n-\frac{I(t)}{F_{\inf}}-\frac
{R(t)}{F_{\inf}}{\large )}\text{ }I(t)-\gamma I(t),\label{eq2}\\
\frac{dR(t)}{dt}  &  =\gamma I(t). \label{eq3}%
\end{align}

But, we can define the new quantities
\begin{align}
\ N &  =n\text{ }F_{\inf}<N_{pop},\\
\beta &  =\frac{p}{F_{\inf}},
\end{align}
where the total population of the country had been called $N_{pop}$ . It is
larger than the number of persons $n$ $F_{\inf}$ living in the $F_{\inf}$
infected families. In terms of these constants the set of equations
(\ref{eq1},\ref{eq2},\ref{eq3}) can be finally rewritten in the typical SIR
form
\begin{align}
\frac{dS^{\ast}(t)}{dt} &  =-\beta\text{ }{\Large (}N-I(t)-R(t){\large )}%
\text{ }I(t),\label{ef1}\\
\frac{dI(t)}{dt} &  =\beta\text{ }{\Large (}N-I(t)-R(t){\large )}\text{
}I(t)-\gamma I(t),\label{ef2}\\
\frac{dR(t)}{dt} &  =\gamma I(t).\label{ef3}%
\end{align}

Therefore, it was argued that a system under stationary isolation condition
should also satisfy the SIR equations when the statistical fluctuations on the
number of infected and recovered cases can be disregarded. This assumption can
be expected to be obeyed for large populations.  This result implies that the
infection of countries under stationary isolation measures can be exactly
described by a SIR model with modified parameters, such that the $N$
 parameter is lower than the total number of habitants in the country.

\subsection{ Retardation effects}

Before adopting the final form of the equations for the model, we estimate
that it should be also taken into account that there is a time interval that
should pass after the infection day, in order that a given person recovers the
healthy state. Clearly, this time is an stochastic quantity which can not be
precisely defined. However, we will attempt to describe this effect by
assuming a definite mean time interval for recovery. It also becomes clear
that for epidemics which appear and disappear within a short time period, this
effect can have a more important influence on the dynamics of the process. In
order to consider this element, we will introduce in the proposed a model a
retardation effect. It will be approximately considered that the rate of
recovering at a given time $t$ is proportional not to the number of infected
persons at this same time, but at a past time $t-\tau.$ The parameter $\tau$
is assumed to be fixed in order to simplify the discussion of the stochastic
nature of the retardation effects.

Finally, the system of equation including these effects will be considered in
the form
\begin{align}
\frac{\partial S^{\ast}(t)}{\partial t}  &  =-\beta\text{ }S^{\ast}(t)\text{
}I(t),\\
\frac{\partial I}{\partial t}  &  =\beta\ \ S^{\ast}(t)\text{ }I(t)-\gamma
\text{ }I(t-\tau),\\
\frac{\partial R}{\partial t}  &  =\gamma\text{ }I(t-\tau),
\end{align}
which retains valid the general conservation relation%
\begin{equation}
N=S^{\ast}(t)+I(t)+R(t).
\end{equation}

It will be helpful to also write the equation in terms of the so called
logistic equation by defining the quantity
\[
S(t)=I(t)+R(t).
\]

Then, the final equations to be further employed take the expression
\begin{align}
\frac{dS(t)}{dt} &  =\beta(N-S(t))I(t),\label{eqr1}\\
\frac{dI}{dt} &  =\beta\ (N-S(t))I(t)-\gamma I(t-\tau),\label{eqr2}\\
\frac{dR}{dt} &  =\gamma I(t-\tau).\label{eqr3}%
\end{align}

\section{Initial conditions and the symptomatic to total cases ratio.}

We will write a set of initial conditions being appropriate to the situation
at any initial moment being after the specific day at which isolation period
is established. Let us call $t_{o}$ a date at which the initial conditions are
imposed. After that, the isolation measures are assumed to be maintained in a
stationary form for the whole population. Assuming numbers of susceptible
$S$, infected $I$ and recovered $R$ persons as functions of time, we can write
for the initial conditions of the set of retarded equations (\ref{eqr1}%
,\ref{eqr2},\ref{eqr3}) for the model
\begin{align}
I(t)  &  =I(t)\text{ \ \ for all \ \ }t\leq t_{o},\label{in1}\\
R(t_{o}^{+})  &  =R(t_{o}^{-}),\label{in2}\\
S(t_{o}^{+})  &  =S(t_{o}^{-}). \label{in3}%
\end{align}

Since the only retarded quantity is the active infected number, the already
measured values of the quantity $I(t)$ at all times $t\leq t_{o}$ are assumed
as initial conditions for the set of equations.

 One important point should be also considered: the ratio $k$ between the
observed number of infected persons and the total number of them is being a
number smaller than unit. We will use for defining it, the evaluation given in
the literature that nearly the 80 \% of the infected persons are asymptomatic
\cite{5} . That is, they show the sickness in a mild form which is not
detected by the Health Systems. Then, we will assume the value%
\begin{equation}
k=0.2,
\end{equation}
for the ratio between the number of detected symptomatic cases and the total
number of cases.

Since the quantities entering the equations (\ref{eqr1},\ref{eqr2},\ref{eqr3})
will be considered defining the total number of confirmed $S$, active $I$ and
recovered $R$ cases, the initial conditions will be defined in the form%
\begin{align}
S(t_{o})  &  =\frac{S_{obs}(t_{o})}{k},\label{inc1}\\
I(t)  &  =\frac{I_{obs}(t)}{k}\text{ \ \ for all \ \ }t\leq t_{o}%
,\label{inc2}\\
R(t_{o})  &  =\frac{R_{obs}(t)}{k}, \label{inc3}%
\end{align}
where $S_{obs}(t)$, $I_{obs}(t)$ and $R_{obs}(t)$ are the observed numbers of
confirmed, active infected and recovered cases, respectively. These numbers
after divided by $k$ give the corresponding numbers of total cases which are
the quantities for which the SIR with retardation equations were derived.

In order to define a smaller range of values for the free parameter $N$ we
will also employ in place of it, the quantity
\begin{equation}
n_{m}=\frac{N}{\frac{I_{obs}(t_{o})}{k}}, \label{n}%
\end{equation}
which is the mean number of infected family members as defined in (\ref{nI}),
after assuming that at the initial time $t_{o}$ any infected family have at
most only one infected person. That is, $F_{inf}=\frac{I_{obs}(t_{o})}{k}$ .
This interpretation seems reasonable at an initial time where the number of
infected persons is very much smaller than the whole population.

In the final section we will solve the set of equations (\ref{eqr1}%
,\ref{eqr2},\ref{eqr3}) in order to check the applicability of the model in
reproducing the \ infection data already measured for various countries. Most
of the countries considered will be ones in which well defined isolation
measures were adopted which were maintained stationary in time.

\section{Small number of cases  solutions:  link with the linear model at
large populations }

In this section we will clarify the  links between  the effective SIR
solutions for epidemics in which  the total number of infected persons is
very much smaller than the total population of the country and the solutions
for SIR for the total population number. When the  number of infected
persons is very much smaller than the population,  the SIR for the total
population reduces to a linear system of differential equations \cite{ours}. But, if the
parameters are constant,  the solutions become exponentials which  decrease, increase
or remain constant all the time. Thus, the only way of describing maxima of
the infection curves is by assuming a time dependence of the parameters.
 However, as it was mentioned before, the recovery rate $\gamma$ is naturally expected to remain
constant, because it should only depend on the standard medical treatments of
the Health System of the population,  in a given country. Thus, the maxima of
the infection  curves for many countries in which the number of cases is very
much lower than the population, must  be necessarily attained by assuming a time dependent
contagion rate $\beta(t).$

Below, it will shown that for each solution of the effective SIR
equations with constant parameters, the time dependent numbers of susceptible, infected and
recovered cases, also exactly satisfy  the SIR for the total population,
but for a time dependent contagion rate $\beta(t)$. The  explicit analytic
formula of $\beta(t)$ is  derived here.

Consider the effective SIR equations with time dependent contagion rate
$\beta(t)$, for the whole population number  (\ref{eqr2})
\begin{equation}
\frac{dI}{dt}=\beta(t)(N-S(t))I(t)-\gamma I(t-\tau),
\end{equation}
in which the parameter $N$ is the  population number of a country,  and
assume that the total number of infected cases $S(t)=I(t)+R(t)$  up to a
given time $t$, satisfies
\[
I(t)+R(t)\ll N.
\]

Then, the SIR equations  reduce to the linear differential equation
\begin{equation}
\frac{dI}{dt}=\beta(t)\text{ }N\text{ }I(t)-\gamma I(t-\tau).\label{SIRPop}%
\end{equation}

\subsection{The effective SIR equations}

Let us now consider a solution of  the effective SIR equations with constant
parameters $(N_{e},\beta_{e})$ written in the form %
\begin{align}
\frac{dS_{e}(t)}{dt} &  =\beta_{e}(N_{e}-S_{e}(t))I_{e}(t),\label{se}\\
\frac{dI_{e}(t)}{dt} &  =\beta_{e}\ (N_{e}-S_{e}(t))I_{e}(t)-\gamma
I_{e}(t-\tau),\label{ie}\\
\frac{dR_{e}(t)}{dt} &  =\gamma I_{e}(t-\tau),\label{re}%
\end{align}
in which the recovery rate $\gamma$ is common with the one in
(\ref{SIRPop}).

Then, if the  following relation is exactly satisfied
\[
\beta(t)\text{ }N=\beta_{e}\ (N_{e}-S_{e}(t)),
\]
the solution ($S_{e}(t)),I_{e}(t),R_{e}(t))$ of the effective SIR equations is
also a solution of linear SIR equations for the total population.  But,
employing  (\ref{ie})  the analytic time dependence of the contagion
parameter takes the form
\begin{equation}
\beta(t)\text{ }=\frac{1}{N}(\frac{1}{I_{e}(t)}\frac{dI_{e}(t)}{dt}%
+\gamma).\label{betat}%
\end{equation}

This time dependence is  furnished by the known  parametric explicit solution
of the SIR equations with constant parameters (See reference \cite{harko}),
which we reproduce below for completeness.  The  number of susceptible
$S_{e}$, infected $I_{e}$ and  $R_{e}$  cases infected and the time $t$, are
defined as  functions of the parametric function  $s_{\eta}(t)$  by the
relations \cite{harko}
\begin{eqnarray}
t  & = & \int_{\exp(-\frac{\beta}{\gamma}n_{3})}^{s_{\eta}(t)}\frac{ds}%
{s(-\beta_{e}\text{ }N_{e}-\gamma\ln(s)+\chi_{0}\beta_{e}\text{ }s)}%
,\label{t}\\
S_{e}(t)  & = & \chi_{0}\text{ }s_\eta(t),\label{s}\\
I_{e}(t)  & = & \frac{\gamma}{\beta_{e}}\ln(s_{\eta}(t))-\chi_{0}\text{ }s_{\eta
}(t)+N_{e}, \label{i}\\
R_{e}(t)  & = & -\frac{\gamma}{\beta_{e}}\ln(s_{\eta}(t)),\label{r}%
\end{eqnarray}
in which  $\eta=(\beta,\gamma,N_{e},n_{1},n_{2}.n_{3})$ is the set of free
parameters  of the solution, and  the initial conditions
had been fixed as
\begin{eqnarray}
S_{e}(0)  & = & n_{1}, \label{n1}\\
I_{e}(0)  & = & n_{2}, \label{n2}\\
R_{e}(0)  & = & n_{3}, \label{n3}\\
N_{e}  & = & n_{1}+n_{2}+n_{3}. \label{Ne}%
\end{eqnarray}

In the above relations the  constants $ \chi_{0}$ is defined by the
relation
\[
\chi_{0}=n_{1}\exp(\frac{\beta}{\gamma}n_{3}).
\]
\begin{figure}[h]
\begin{center}
\includegraphics[width=.6\textwidth]{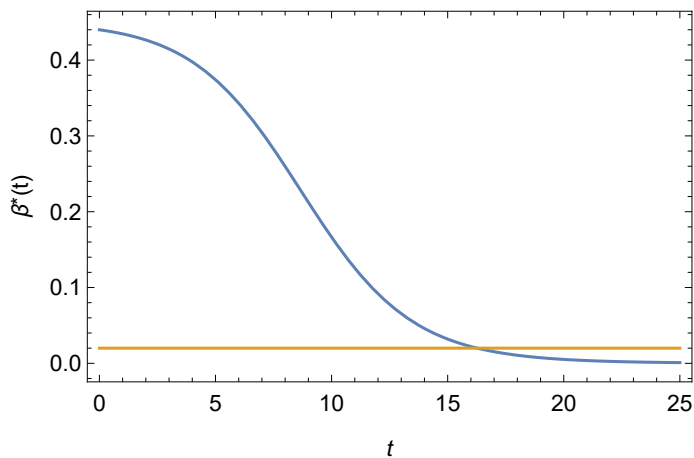}
\end{center}
\caption{The curve gives the time dependence of the  contagion parameter $\beta(t)$. This function after substituted in the linear equations for the SIR for total population,  allows that the solutions of the effective SIR equations for the same parameter values also satisfy them.  The function depends on the parameters and initial conditions
 of the effective SIR equations. These values can be varied in order to find the specific time dependence which closely fits the data for the susceptible, infected and recovered cases at the beginning  of the infection. Then, for future times, the solution can be compared with the data to be taken.  }%
\label{beta}%
\end{figure}

Making use of these definitions, the  expression for the time dependent
contagion parameter (\ref{betat})
and the function $R_0(t)$ may be written as
\begin{align}
N\beta(t)  & =\beta^{\ast}(t)=\beta_{e}n_{1}\exp(\frac{\beta_{e}}{\gamma}%
n_{3})\text{ }s_{\eta}(t),\\
R_{0}(t)  & =\frac{\beta^{\ast}(t)}{\gamma}=\frac{1}{\gamma}\frac{1}{I_{e}(t)}\frac{dI_{e}(t)}{dt}+1 \nonumber\\
          & =\frac{1}{\gamma}\beta_{e}n_{1}\exp(\frac{\beta_{e}}{\gamma}%
n_{3})\text{ }s_{\eta}(t).
\end{align}
The figure \ref{beta} shows the time dependent $\beta(t)$ for which the effective SIR solutions associated to the specific parameter values
 \begin{eqnarray}
\eta & = & (\beta,\gamma,N_{e},n_{1},n_{2}.n_{3}) \nonumber \\
& = & (0.01,0.02,45,44,1,0). \label{data}
\end{eqnarray}
  The horizontal line plots the constant value of $\gamma$ and the intersection point  defines
 the maximum of the curve of active infected cases.

\subsection{  Proposals for predictive time dependence methods  for  epidemics  }

\begin{figure}[h]
\begin{center}
\includegraphics[width=.6\textwidth]{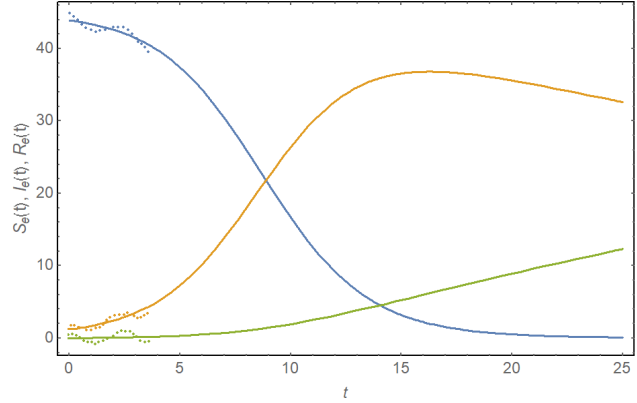}
\end{center}
\caption{  The solution for susceptible $S_e(t)$, infected $I_e(t)$ and recovered $R_e(t)$ cases  for the effective SIR model. They are  associated to the parameters chosen in (\ref{data}) for evaluating the
$\beta^*(t)$. The point plots near the initial time $t=0$  are only auxiliary illustrations representing the observed data for each of the three quantities $S_e(t)$, $I_e(t)$ and $R_e(t)$. }%
\label{figModf1}%
\end{figure}

The results of this section can be used for attempting to predict the
 epidemics in the cases that rigorous stationary isolation measures  had
been imposed in the Society.  The method only will be  sketched here, but it
will be investigated in more detail elsewhere.  Under  the stationary
assumption, we had argued in the first Section  that a SIR model should define
the time evolution of the infection.  The method can be easily
described by examining figure  \ref{figModf1}. It  plots the SIR model
solution (\ref{n1},\ref{n2},\ref{n3}) for the particular set of parameters and initial
conditions
\begin{align}
\eta & =(\beta,\gamma,N_{e},n_{1},n_{2}.n_{3}) \nonumber\\
& =(0.01,0.02,45,44,1,0).
\end{align}

The figure shows the susceptible $S_{e}(t)$, infected $I_{e},$ and recovered
$R_{e}$ number of particles solving the effective SIR equations.

Let us assume now that the Health systems had observed the three types of
numbers during a given time interval at the beginning of the epidemics.  These
measured numbers are symbolized by the irregularly distributed points being close to
 the three continuous solutions of the model.  Then, the discussion
in this paper allows  shows that for describing the  epidemics under
stationary isolation measures, it is possible two use two methods. Let us
describe them below.

\subsubsection{a) Solve the effective SIR equations by matching the data
near $t=0$ }

1) Firstly,  fix the  there initial conditions $(n_{1},n_{2},n_{3})$ as
 approximately defined by the values observed by the Health System at the
origin of time $t=0.$

2) Next, evaluate the solutions (\ref{t})-(\ref{r}) by varying the
values  of the resting parameters $(\beta,\gamma,N_{e})$ in order to
 make  the best possible fitting of the data in the neighborhood of  the
initial time $t=0$.  If the system is really stationary, the data should
be followed ahead in the time by the solution. If not, a deviation caused by
the  non stationary isolation measures can appear.

 It should be remarked, that the assumption about that a SIR model
with an adjustable population parameter should describe the infection
evolutions had been employed in the literature. See by example the recent
reference \cite{DamianNana}. Therefore, the mean value of the argue in Section
1 is a theoretical checking that a SIR model with adjustable
 population number, and  contagion and recovering parameters should describe
the evolution of the susceptible, infection and recovering numbers in the
 epidemics.  That is,  to establish that the solution of the effective
SIR model should be expected to describe  the evolution of the infection
assumed that the  isolation measures are really stationary.

\subsubsection{b) Solve the  linear SIR equation for the whole population
under a  time varying effective contagion parameter }

 1) In this case, the first step is to consider the linear SIR equations in
which  $N$ is the whole population and the number of infected persons is very
much small than $N$:
\begin{align}
\frac{dI}{dt}  & =(\beta^{\ast}(t)-\gamma)I(t),\\
I+R  & \ll N,
\end{align}
we have considered zero retardation $\tau=0$ for simplicity,  but retardation
can be included. This equation is equivalent to the effective SIR ones, if the
contagion parameter $\beta^{\ast}(t)$ is defined in terms of the solutions
of the effective SIR (\ref{t},\ref{s},\ref{i},\ref{r}) as
\[
\beta^{\ast}(t)=\beta_{e}n_{1}\exp(\frac{\beta_{e}}{\gamma}n_{3})s_{\eta}(t).
\]

2) Therefore, again varying the parameters $\eta=(\beta_{e},\gamma,N_{e}%
,n_{1},n_{2}.n_{3})$  to attain the best fit with observed infection data near the beginning of the infection,
the solution should give the same effective SIR  solutions for any further time, if the isolation measures are stationary.

 It should be remarked that the discussion done here clarifies the links
between  the solutions sought by  assuming the SIR for the whole
population and the ones obtained by optimizing  an effective SIR as a
function of the parameter. It is clear that after assuming the whole
population, a  time dependence  of the contagion parameter should exist in
order to  reproduce  contagion evolutions  showing maxima in the evolution.
Such maxima can not exist under a simple exponential  time dependence.

\section{Application to epidemic data in various countries}

In this section, we will present the result of the solutions of the model in
describing the infection data associated to the following countries Iceland,
New Zealand, South Korea, Cuba, United States and Mexico. The first four are examples
of counties in which isolation measures restraining the contact between
persons in the population were imposed in a stationary form, and were
maintained during large time intervals. They were selected precisely by these
properties of the mentioned confinement rules, which are appropriate for
checking the validity of modified SIR equations. Finally, the cases of United
States and Mexico were also studied, for which it can not be considered that stationary
confinement measures had been imposed. However, these countries show
radical development of the sickness at present (9 June 2020). Therefore, it becomes of interest
to estimate rough predictions of the model for the occurrence of the peak and
the maximal number of active cases.

Below, a subsection will be devoted to apply the model to each of these countries.

\subsection{Iceland}

The data for the daily number of confirmed, active infected and recovered
cases for all the countries considered in this work were obtained from
reference \cite{9}. The number of the days in the time axes of the figures
appearing below are defining the time lapse passed from the 22 January 2020 at
which reference \cite{9} started  to publish the data. For Iceland the time
for the initial condition was chosen as
\begin{equation}
t_{o}=60.
\end{equation}

\begin{figure}[h]
\begin{center}
\includegraphics[width=.6\textwidth]{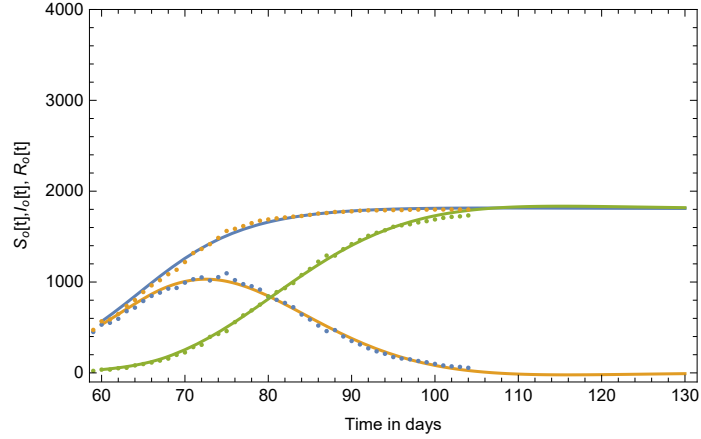}
\end{center}
\caption{ The plots show the comparison between the solutions of the SIR
\ equations (continuous plots) and the infection data (point like plots) \ for
Iceland. The blue curve shows the confirmed cases, the red one gives the
active ones and the green line define the recovered patients. Note that a
solution of the SIR model with retardation closely approaches the observed
data, for the properly chosen values of the four parameters $N,\beta,\gamma$
and $\tau.$ The non vanishing value of $\tau$ indicates that retardation plays
a role in the description of the time evolution. process. }%
\label{fig1f}%
\end{figure}

The general initial conditions defined in (\ref{inc1},\ref{inc2},\ref{inc3})
were specified in the form
\begin{align}
S(t_{o})  &  =\frac{S_{obs}(t_{o})}{k}=2840,\\
I(t)  &  =\frac{I_{obs}(t)}{k}\text{ =}13.9888\exp(0.1731(t-39))\ \text{\ for
all \ \ }t\leq t_{o},\\
R(t_{o})  &  =\frac{R_{obs}(t)}{k}=180,
\end{align}
where $S_{obs}(t_{o}),I_{obs}(t)$ and $R_{obs}(t_{o})$ are the data for the
confirmed, active infected and recovered cases for Iceland in reference
\cite{9} at the day $t_{o}$ and before. The function defining the values of
the infection curve at all the times before $t_{o}$, is obtained by fitting an
exponential behavior to the data for $I_{obs}(t)$ in reference \cite{9}, for
the considered country. In this region of small number of infections, the
exponential solutions are exact and gives an accurate fitting of the initial data.

After fixing the initial conditions, it remains the four free parameters
$\ (n,\beta,\gamma,\tau)$ to be determined for defining the solution which
better can match the data for all the future times $t\geq t_{o}.$ Then, for
the already defined initial conditions we solved the set of equations
(\ref{eqr1},\ref{eqr2},\ref{eqr3}) by attempting to approach the observed data
for Iceland in the best way possible. The figure \ref{fig1f} shows the result
of this process. The continuous curves represent the best exact solutions of
the equations attained for three relevant quantities. The blue curve is the
number of the observed confirmed, the red line shows the observed active cases
and lastly, the  green plot shows the observed recovered cases. The plotted
quantities are the predictions for the "observed" quantities are defined as
\begin{align}
S_{o}(t)  &  =k\text{ }S(t),\\
I_{o}(t)  &  =k\text{ }I(t),\\
R_{o}(t)  &  =k\text{ }R(t),
\end{align}
that is, the total numbers $S(t),$ $I(t)$ and $R$ after multiplied by
$\ k=0.2.$ The curves of points show the corresponding data $S_{obs}%
(t),I_{obs}(t)$ and $R_{obs}(t)$ for Iceland obtained from reference \cite{9}.
The resulting optimal parameter values determining the shown approach of the
solution to the data were:%
\begin{align}
N  &  =9797,\\
\beta &  =172.5\times10^{-7},\\
\gamma &  =0.063,\\
\tau &  =8.
\end{align}
The figure \ref{fig1f} clearly indicates that a solution of a SIR model exists
which reproduces in a satisfactory form the observed epidemic data curves for
Iceland. It should be noted that it was attempted to match the observed and
solved time evolution by avoiding to introduce retardation effects, that is
fixing $\tau=0.$ However, it turned out to be impossible for us to attain a
similar likelihood as the one shown in figure \ref{fig1f}. Only after choosing
a non vanishing retardation value $\tau$, it became possible to match the two
behaviors as in the figure \ref{fig1f}.

\subsection{New Zealand}

The behavior of the epidemics for New Zealand was very similar to the one of
Iceland. After equally getting the initial data from reference \cite{9} and
choosing the time for the initial condition to be
\begin{equation}
t_{o}=64,
\end{equation}

\begin{figure}[h]
\begin{center}
\includegraphics[width=.6\textwidth]{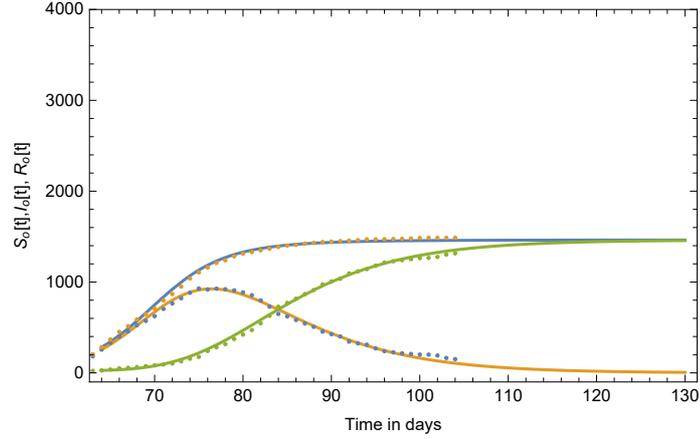}
\end{center}
\caption{ The figure presents the confirmed, active and recovered cases as
continuous blue, red and green plots, respectively. It can be noted that in
this case of New Zealand the infection data (point like plots) the solution of
the SIR model with retardation, also well resembles the observed data, when
proper values of the four parameters $N,\beta,\gamma$ and $\tau$ are chosen$.$
Observe that retardation is also important in describing the time evolution. }%
\label{fig2f}%
\end{figure}the initial conditions (\ref{inc1},\ref{inc2},\ref{inc3}) were
written for this country as follows%
\begin{align}
S(t_{o})  &  =\frac{S_{obs}(t_{o})}{k}=1415,\\
I(t)  &  =\frac{I_{obs}(t)}{k}\text{ =}0.0885\exp(0.3187(t-39))\ \text{\ for
all \ \ }t\leq t_{o},\\
R(t_{o})  &  =\frac{R_{obs}(t)}{k}=135,
\end{align}
where $S_{obs}(t_{o}),I_{obs}(t_{o})$ and $R_{obs}(t_{o})$ are again the data
for the confirmed, active infected and recovered cases for New Zealand found
in reference \cite{9} for the time $t_{o}$ and before. As for the previous case
the function defining the values of the infection curve at all times before
$t_{o}$, is evaluated after fitting an exponential behavior to the data for
$I_{obs}(t)$ in reference \cite{9}.

Then, in a closely form as in the previous subsection we solved the set of
equations (\ref{eqr1},\ref{eqr2},\ref{eqr3}) successively in order to approach
the observed data in this case for New Zealand. The figure \ref{fig2f} shows
the outcome. As before, the continuous curves represent the best exact
solutions of the equations attained for the three relevant quantities. The
blue curve again is the number of the observed confirmed, the red one depicts
the observed active cases and the green curve illustrates the observed
recovered cases: defined as before by
\begin{align}
S_{o}(t)  &  =k\text{ }S(t),\\
I_{o}(t)  &  =k\text{ }I(t),\\
R_{o}(t)  &  =k\text{ }R(t).
\end{align}
The curves of points again show the corresponding data $S_{obs}(t),I_{obs}(t)$
and $R_{obs}(t)$ in this case for New Zealand, also obtained from reference
\cite{9}. The resulting parameter values for which maximal likelihood was attained took
the values
\begin{align}
N  &  =7424,\\
\beta &  =360.5\times10^{-7},\\
\gamma &  =0.063,\\
\tau &  =5.5.
\end{align}

The figure \ref{fig2f} again evidence that a solutions of a SIR model
appropriately reproduce the form the data curves also in this case for New Zealand.

It also was attempted to match the observed and evaluated time
evolution without considering retardation ($\tau=0$).  As before, it was not
possible to fit the data as closely as in figure \ref{fig2f}.  A retardation
value $\tau=5.5$ days was required to adjust the curves.

\subsection{South Korea}

The initial conditions for the study of the South Korea epidemic data were
fixed at
\begin{equation}
t_{o}=39,
\end{equation}
and the chosen starting and retarded values for the total $S,$ $I$ and $R$
defined as follows
\begin{align}
S(t_{o})  &  =\frac{S_{obs}(t_{o})}{k}=53825,\\
I(t)  &  =\frac{15.1135\exp(0.2803(t-20))\text{\ }}{k}\text{for all \ \ }t\leq
t_{o},\\
R(t_{o})  &  =\frac{R_{obs}(t)}{k}=46530.
\end{align}
\begin{figure}[h]
\begin{center}
\includegraphics[width=.6\textwidth]{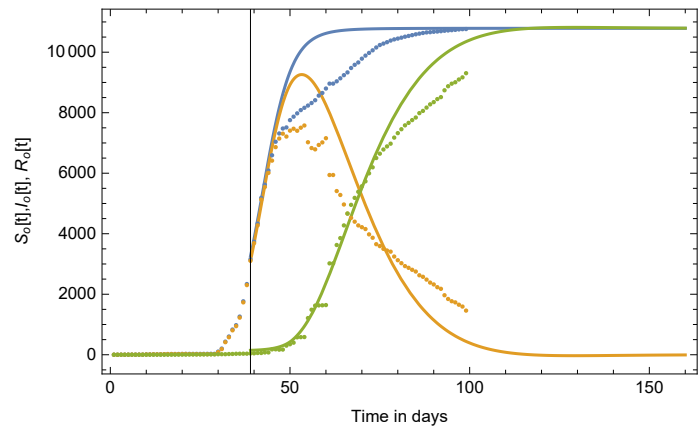}
\end{center}
\caption{ The figure expose the results for the infection data corresponding
to South Korea. As before, the blue, red and green continuous curves show the
solutions for the number of the observed confirmed $S_{o}$, active $I_{o}$ and
recovered $R_{o}$ cases, respectively. Up to the day 43 (near 4 March 2020) the
solutions again well match the observed data. After this time the continuous
curves deviate from the observations. However, precisely the March 4, South
Korea appreciably increased the isolation measures, breaking in this way the
stationary character of the imposed isolation rules. Thus, the figure, also
confirm the conclusion about that a modified SIR models describe the
infections under stationary isolation rules. }%
\label{fig3f}%
\end{figure}

After following exactly the same steps as in the discussion in the previous
two subsections, the values of the four free parameters in order to
approach the data as shown in figure \ref{fig3f}, were found  in the form
\begin{align}
N  &  =398305,\\
\beta &  =47.1645\times10^{-7},\\
\gamma &  =0.035,\\
\tau &  =12.
\end{align}

The analysis brought an additional new element to the former discussions in
previous subsections. As shown in figure \ref{fig3f} the three solutions of the
equations, are shown by continuous curves. The blue one is the observed number
of confirmed cases, the red one corresponds to the observed number of active
cases and lastly, the green one presents the number of recovered cases.

However, near the day 43 (4 March 2020) it can be noted that the curve for the
solution of the SIR model starts deviating from the observed data. This
observation led us to search in the literature information about how were
implemented the isolation measures in South Korea. Surprisingly, it was found
that precisely the 4 March, the country had defined a drastic increase in the
rigor of such measures. Therefore, the mentioned deviation of the
solutions from the observed data, can be interpreted as produced by the
breaking of the stationary character of the confinement rules, which were
established up to such a moment. Thus, the example of South Korea can be also
supporting the main conclusions of the present work.

\subsection{ Cuba}

In the case of Cuba, in which tight isolation measures were also taken, they
were established nearly the date 03.24.20, that is, at the time measured in
days from the 01.22.20
\begin{equation}
t_{o}=63.
\end{equation}
 The initial conditions chosen for the equations (\ref{eqr1},\ref{eqr2}%
,\ref{eqr3}) at that moment were
\begin{align}
S(t_{o})  &  =\frac{S_{obs}(t_{o})}{k}=240,\\
I(t)  &  =\frac{I_{obs}(t)}{k}\text{ =}\frac{0.0303054\exp(0.305211(t-39))}%
{k}\text{\ for all \ \ }t\leq t_{o},\\
R(t_{o})  &  =\frac{R_{obs}(t)}{k}=5\text{ }.
\end{align}

\begin{figure}[h]
\begin{center}
\includegraphics[width=.6\textwidth]{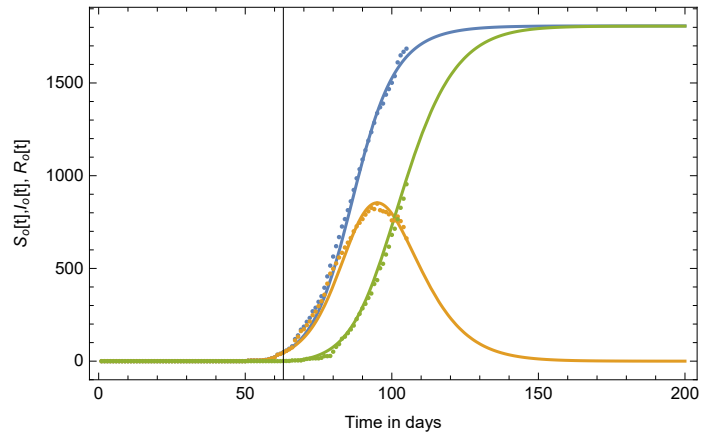}
\end{center}
\caption{ The blue, red and green continuous curves present the Cuba's
solutions for the observed confirmed $S_{o}$, active $I_{o}$ and recovered
$R_{o}$ cases, respectively. The stationary character of the isolation
measures sustained in the country, is consistent with  appreciable degree of
matching between the solution and the observations shown by the figure. }%
\label{fig4f}%
\end{figure}

As in the previous subsections, the function $I_{obs}(t)$ defines the number of observed
infections for all times before$\ t_{o}$, information which is needed to
determine the solution of the retarded equations. In the present case it was
evaluated in a continuous way by interpolating the observed discrete values of
the infection curve before the day $63$. Reiterating the process of
determining the best parameters for which the solution of the SIR equations
with retardation shows a close likelihood with the data, the attained  parameter
values resulted in
\begin{align}
N  &  =9775,\\
\beta &  =170.9\times10^{-7},\\
\gamma &  =0.06,\\
\tau &  =6.5\text{ }.
\end{align}

The figure \ref{fig4f} illustrates the observed confirmed cases $S_{o}$
$(t)=k$ $S(t)$ with the blue curve, the active ones $I_{o}(t)=k$ $I(t)$ \ with
the red one and the recovered cases $R_{o}(t)=k$ $R(t)$ are shown in the green
curve. The results further support that the solutions of a modified SIR model
well describe the dynamics of the covid-19 epidemics under stationary
isolation measures.

\subsection{United States}

We also have considered the case of United States. This country can not be
considered as one in which rigorous and stationary in time isolation
conditions had been implemented to stop the epidemic.
\begin{figure}[h]
\begin{center}
\includegraphics[width=.6\textwidth]{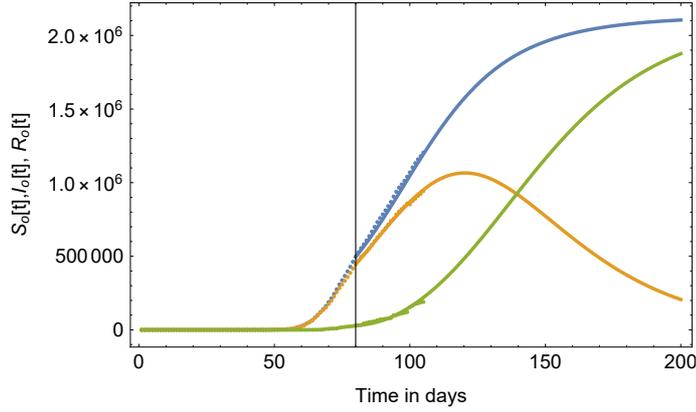}
\end{center}
\caption{ The figure shows the solutions of the SIR model with retardation for
the case of United States. Again the continuous curves present the confirmed
$S_{o}$, active $I_{o}$ and recovered $R_{o}$ cases respectively. For US the
stationary character of the isolation measures can not be assured. However,
due to the strong character of the infection in this country the SIR solutions
were compared with the observed data, in order to estimate important
quantities as the date and amount of infected cases in the peak of the
infection curve. The figure shows a reasonable degree of matching of a SIR
solution with the observed data. The resulting estimation for the date of the
peak is May 20, 2020 with a maximum of nearly $1.064$\ $\times10^{6}$ observed
active cases. }%
\label{fig5f}%
\end{figure}
However, we consider the application of the model by two main
reasons: the country's data show a somewhat regular evolution which seemingly
could be approximately approached by a solution of the SIR equations. In
second place, due to the strong behavior of the epidemic in the country, it
looks helpful to explore the predictions of the model in order to estimate the
date for the occurrence and number of infected cases of the maximum of the
infected curve.

The initial time for searching a solution of the equations (\ref{eqr1}%
,\ref{eqr2},\ref{eqr3}) was
\begin{equation}
t_{o}=80,
\end{equation}
and the initial (for $S$ and $R$ ) and retarded data (for $I$) in this case
were
\begin{align}
S(t_{o})  &  =\frac{S_{obs}(t_{o})}{k}=2.48268\times10^6,\\
I(t)  &  =\frac{f_{int}(t)}{k}\text{ for all \ }t\leq t_{o},\\
R(t_{o})  &  =\frac{R_{obs}(t_{o})}{k}=143950,
\end{align}
in which $\frac{S_{obs}(t_{o})}{k}$ and $\frac{R_{obs}(t_{o})}{k}$ are the
number of observed infected and recovered cases at the day $t_{o}=80$ after
divided by $k,$ in order to define the total number of cases at this time. The
function $f_{int}(t)$ defines the retarded data for the number of active cases
as a continuous interpolation function of the set of discrete data points
$I_{obs}$ for the observed active cases. The number of observed three types of
cases were taken from reference \cite{9}. As in all the previous cases, the
search for solutions closely representing the data furnished the following set
of parameters
\begin{align}
N  &  =1.1358\times 10^7,\\
\beta &  =0.056\text{ }\times\text{\ }10^{-7},\\
\gamma &  =0.022,\\
\tau &  =16.
\end{align}

The figure \ref{fig5f} shows the solution (continuous lines) in comparison
with the data (point like curves). As always, the plots correspond to the
observed number of confirmed (in blue line), infected (red line) and recovered
(green line) cases $S_{o}(t),$ $I_{o}(t)$ and $R_{o}(t)$, which as before, are
the total numbers (which the solution furnishes) after multiplied by $k$, that
is
\begin{align}
S_{o}(t)  &  =k\text{ }S(t),\\
I_{o}(t)  &  =k\text{ }I(t),\\
R_{o}(t)  &  =k\text{ }R(t).
\end{align}

The figure \ref{fig5f} shows that a definite solution of a modified SIR model
reasonably match the observed data for the three quantities. Only the
resulting value for the recovery parameter $\gamma$ took a slightly small
value nearly a half of the values for the previously discussed countries. In
addition, the probably lack of stationary character of the isolation measures,
makes clear that assuming that they are stationary along the country is a very
strong assumption. Thus, the maximum of the active cases and the date at which
it appears can be only taken as a estimate of those quantities. The predicted
values are
\begin{align*}
t_{\max}(US)  &  =120\equiv\text{ May 20,}\\
I_{\max}  &  =1.064\text{ \ }10^{6}.
\end{align*}

\subsection{Mexico}

Finally, we discuss the case of  Mexico. In similar way as USA, in Mexico
rigorous and stationary in time isolation conditions had not been imposed.
\begin{figure}[h]
\begin{center}
\includegraphics[width=.6\textwidth]{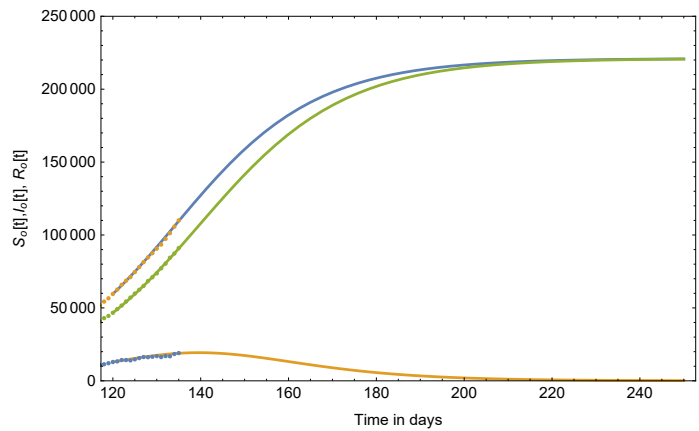}
\end{center}
\caption{ The figure shows the solutions of the SIR model with retardation for
the case of  Mexico. As before, the continuous curves present the confirmed
$S_{o}$, active $I_{o}$ and recovered $R_{o}$ cases respectively. For Mexico,
also, the stationary character of the isolation measures can not be assured.
However, due to the large population of the country and the irregularities in
the evolution of the infection the SIR solutions were compared with the
observed data, seeking to estimate important quantities as the date and amount
of infected cases in the peak of the infection curve. In spite of the above
remarks, the figure shows a reasonable degree of matching of a SIR solution
with the observed data. The resulting estimation for the date of the peak is
June 10, 2020 with a maximum of nearly $20,000$ observed active cases. }%
\label{fig6f}%
\end{figure}
As for the case of USA we consider the application  of the model
 because the country is a large one and  the infection data had showed an
irregular evolution.

The initial time for searching a solution of the equations (\ref{eqr1}%
,\ref{eqr2},\ref{eqr3}) was
\begin{equation}
t_{o}=120,
\end{equation}
and the initial (for $S$ and $R$ ) and retarded data (for $I$) in this case
were
\begin{align}
S(t_{o})  &  =\frac{S_{obs}(t_{o})}{k}=297835,\\
I(t)  &  =\frac{f_{int}(t)}{k}\text{ for all \ }t\leq t_{o},\\
R(t_{o})  &  =\frac{R_{obs}(t_{o})}{k}=233310,
\end{align}
in which as in the previous subsections $\frac{S_{obs}(t_{o})}{k}$ and
$\frac{R_{obs}(t_{o})}{k}$ are the number of observed infected and recovered
cases at the day $t_{o}=80$ after divided by $k,$ in order to define the total
number of cases at this time. The function $f_{int}(t)$ defines the retarded
data for the number of active cases as a continuous interpolation function of
the set of discrete data points $I_{obs}$ for the observed active cases.
Again, the number of observed three types of cases were taken from reference
\cite{9}. For Mexico the search for solutions closely representing the data
gave the following results
\begin{align}
N  &  =220,000,\\
\beta &  =1.20\times\text{\ }10^{-7},\\
\gamma &  =0.18,\\
\tau &  =0.
\end{align}

Figure \ref{fig6f} exhibits the solution (continuous lines) in comparison with
the data (point like curves). The plots correspond to the observed number of
confirmed (in blue line), infected (red line) and recovered (green line) cases
$S_{o}(t),$ $I_{o}(t)$ and $R_{o}(t)$.  These quantities, as before, are the
total numbers (which the solution furnishes) after multiplied by $k$, that is
\begin{align}
S_{o}(t)  &  =k\text{ }S(t),\\
I_{o}(t)  &  =k\text{ }I(t),\\
R_{o}(t)  &  =k\text{ }R(t).
\end{align}

The figure \ref{fig6f}  indicates that a solution of a modified SIR model
reasonably match the observed data for the three quantities. It is of interest
 to note that the recovery parameter $\gamma$  has
a slightly high value, being larger than the ones for the all previously
discussed countries. It is curious that the number of recovered cases
in Mexico is from the start of the epidemic,  larger that
 the number of active cases. This property is seemingly compatible  with
the observed high value of the recovery parameter.

Like for U.S.A., the probably lack of stationary character of the isolation
measures, suggests doubts about the validity of a SIR model with constant
parameters, as the one applied. Thus, the maximum of the active cases and the
date at which it appears can be only taken as a rough  estimate of those quantities.
The resulting   values were
\begin{align*}
t_{\max}(Mexico)  &  =140\equiv\text{ June 10,}\\
I_{\max}  &  =19,300.
\end{align*}

\section*{Conclusions}

It is argued that under stationary in time isolation measures, the SIR
equations should be valid for the description of the epidemics, after
disregarding statistical fluctuations. The resulting modified SIR equations
have a reduced population parameter $N$ and a constant  contagion rate $\beta$.
The relevance of the retardation for properly describing the observed
infection curves is also identified. A modified SIR model including
retardation is introduced and applied to describe the infection data for
various countries. The results properly confirm the idea about that when the
isolation measures are stationary in time, the SIR equations solutions
properly describe the observed confirmed, infection and recovering cases daily data.

The results also show that  the solutions of the effective SIR with
adjustable parameters (including the population number) when the total number
of infected cases is very much smaller than the population, also exactly solve
the SIR equations for the whole population.  This  conclusion  allows to solve an existing
 challenging problem about how to define the required time dependence of the contagion parameter when
 the SIR model is decided to be applied to the whole population \cite{14,15,16,17,18,19}.
 The time dependence to be imposed for  describing population under stationary isolation regimes is
 here exactly determined thanks to the also known exact solution of the SIR model \cite{harko}.

 The discussion also allows to propose here two procedures  for predicting the infection curves, after observing the data, up to a time $T$ before the arrival to a maximum number of infected cases. A first  possible
method, is to minimize the square deviation between the solution and the
observed data for all times before the moment $T$. The expectation is that
this minimization process could be able to define the four relevant parameters
of the SIR model with sufficiently small error,  to efficiently predict the
future evolution of the infection, after the chosen time $T$. Another way is
to solve the linear equations for the SIR for the whole population employing the
derived formula for the  time dependent contagion parameter $\beta(t).$  In this case, an analogous
to the above described method of minimizing the square deviation  can be
also employed.

 Finally, the effective SIR model is applied to various countries in which proper isolation measures
had been taken in combating the epidemic. The results for them appropriately
match the observations after properly accounting  for the delay between
recovery and infection processes.  There are also discussed two countries in
which isolation measures can not be assured that had been constantly in time
adopted: USA and Mexico. However, the results for the observations for a not
too large time interval can be also reasonably reproduced.

\section*{Acknowledgements}

We acknowledge the researchers Paulina Ilmonen,  Juan Barranco, Argelia Bernal, Alma
Gonz\'{a}lez, Oscar Loaiza, Dami\'{a}n Mayorga Pe\~{n}a, Octavio Obreg\'{o}n,
Luis Ure\~{n}a, by helpful exchanges during the developing of the work.
NCB also acknowledge the "Proyecto CONACyT A1-S- 37752" and "Laboratorio de
Datos, DCI, de la Universidad de Guanajuato". ACM in addition  wish to acknowledge the
support received form the Network 09, of the Office of Externals Affairs (OEA)
of the International Centre for Theoretical Physics (ICTP) in Trieste, Italy.


\begin{thebibliography}{99}                                                                                               %


\bibitem {1}\textit{The SARS-CoV-2 in Mexico: analysis of plausible scenarios of
behavioral change and outbreak containment}, M. A. Acu\~na-Zegarra et
al.  \url{https://www.medrxiv.org}, 2020.

\bibitem {2}\textit{An updated estimation of the risk of transmision of the
novelcoronavirus (2019-nCov)}, B. Tang, N. L.  Bragazzi, Q. Li, S. Tang, Y Xiao and J. Wu.
\textbf{Infectious Diseases Modelling 5}, 248-255, 2020.

\bibitem {3}\textit{Modelo de infectados para el Edo. de Guanajuato COVID-19} ,
J. Barranco y A. Bernal, unpublished.

\bibitem {4}\textit{Como us\'e las matem\'aticas para predecir
COVID-19 en el estado de Guanajuato, M\'exico}, Lisa Shiller.\newline%
\url{https://www.lisashiller.com/blog-1/cmo-us-las-matemticas-para-} \url{predecir-covid-19-en-el-estado-de-guanajuato-mxico}.

\bibitem {5}\textit{Substantial undocumented infection facilitates the rapid
dissemination of novel coronavirus (SARS-CoV2)}, R. Li, S. Pei, B. Chen, Y. Song,T. Zhang,
W. Yang and  J. Shaman,  \textbf{Science 368}, 489-493, 2020.


\bibitem {6}\textit{Public Health responses to COVID-19 outbreaks on cruise ships
euro Worldwide}, February href euro, March 2020, Moriarty et al.  \newline\url{https://www.cdc.gov/mmwr/volumes/69/wr/mm6912e3.htm}

\bibitem {7}\textit{A SIR epidemic model with time delay and general nonlinear
incidence rate}, M. Li and X. Liu, \textbf{Hindawi Publishing Corporation, Abstract and Applied Analysis Vol. 2014}, Article ID 131257.

\bibitem {8}\textit{Infectious disease models with time-varying parameters and
general nonlinear incidence rate}, X. Liu and P. Stechlinski, \textbf{Applied Mathematical Modelling 36}, 1974-1994, 2012.


\bibitem {9} Website: \url{https://data.humdata.org/dataset/novel-coronavirus-2019-ncov-cases}

\bibitem {10} Website:
\url{https://medium.com/data-for-science/epidemic-modeling-101-or-}\newline \url{why-your-covid19-exponential-fits-are-wrong-97aa50c55f8}

\bibitem {11}Website: \url{https://www.worldometer.info}.

\bibitem {ours}\textit{Modelos SIR modificados para la evolucion del COVID19},  N. G. Cabo-Bizet and A. Cabo Montes de Oca. Submitted and accepted to be published in \textbf{Revista Cubana de Matematicas} in 2020.
\textbf{arXiv:2004.11352v1 [q-bio.PE]} 23 April 2020.

\bibitem {12}\textit{Estimating the number of infections and the impact of non-
pharmaceutical interventions on COVID-19 in 11 European countries}, S. Flaxman, S. Mishra and A. Gandy and S- Bhatt,
\textbf{Nature} \url{doi.org./10.1038/s41586-020-24025-7}, 2020.

\bibitem {13}\textit{Das mysterium um die ansteckungsrate}, Katherine Rydlink,
\textbf{Spiegel Online}, 17 April 2020, \url{https://www.spiegel.de}
\bibitem {DamianNana} \textit{Time-dependent and time-independent SIR models applied to the COVID-19 outbreak in Argentina, Brazil, Colombia, Mexico and South Africa}, N. G. Cabo-Bizet and D. Mayorga Pena,  \textbf{arXiv:2006.12479v1 [q-bio.PE]}, 2020.
\bibitem {harko}\textit{Exact analytical solutions of the Susceptible-Infected-Recovered
(SIR) epidemic model and of the SIR model with equal death and birth rates},  T. Harko, F. S. N. Lobo  and M. K. Mak, \textbf{Applied Mathematics and Computation 236}, 184-194, 2014.
\bibitem {14}\textit{Modeling the Transmission of Middle East Respiratory Syndrome CoronaVirus in the Republic of Korea}Z-Q Xia, J. Zhang, Y-K. Xue, G-Q. Sun, Z. Jin, \textbf{PLoSONE 10},(12): e0144778, 2015.
\bibitem {15}\textit{Predicting the evolution of SARS-COVID-2 in Portugal using an adapted SIR model previously used in South Korea for MERS outbreak},  P. Teles,  \textbf{arXiv:2003.10047v2 [q-bio.PE]} 9 Apr 2020.
\bibitem {16} \textit{Estimation of timevarying reproduction numbers underlying
epidemiological processes: A new statistical tool for the COVID-19 pandemic}, H. G. Hong and Y. Li,  \textbf{PLoS ONE 15},(7): e0236464, 2020.
\bibitem {17}\textit{A Time-dependent SIR model for COVID-19 with Undetectable Infected Persons},
Yi-Cheng Chen, Ping-En Luy, Cheng-Shang Changz and
Tzu-Hsuan Liux, \textbf{arXiv:2003.00122v6 [q-bio.PE]} 28 Apr 2020.
\bibitem {18}\textit{SIR model with time dependent infectivity parameter: approximating the epidemic attractor and the importance of the initial phase}, S. Boatto, C. Bonnet, B. Cazelles and F. Mazenc,   \textbf{hal-01677886}, 2018.
\bibitem {19} \textit{An epidemiological forecast model and software assessing
interventions on COVID-19 epidemic in China}, L. Wang, Y. Zhou, J. He, B. Zhu, F. Wang, L. Tang,
M. Eisenberg and P. X. K. Song,  \textbf{medRxiv preprint} doi: https://doi.org/10.1101/2020.02.29.20029421, posted March 3, 2020.


\end{thebibliography}
\end{document}